\documentclass{nature}

\usepackage{graphicx,amsmath}
\usepackage{booktabs}
\usepackage[flushleft]{threeparttable}

\usepackage{siunitx}
\title{Evaporation-triggered microdroplet nucleation and the four life phases of an evaporating Ouzo drop}

\author{Huanshu Tan$^{1}$, Christian Diddens$^2$, Pengyu Lv$^{1}$, J. G. M.  Kuerten$^{2,3}$, Xuehua Zhang$^{4}$, Detlef Lohse$^{1,5}$}

\begin{document}

\maketitle

\begin{affiliations}
 \item Physics of Fluids group, Department of Science and Technology, Mesa+ Institute, and  J. M. Burgers Centre for Fluid Dynamics, University of Twente, P.O. Box 217, 7500 AE Enschede, The Netherlands,
 \item Department of Mechanical Engineering, Eindhoven University of Technology, P.O. Box 513, 5600 MB Eindhoven, The Netherlands,
 \item Faculty EEMCS, University of Twente, P.O. Box 217, 7500 AE Enschede, The Netherlands,
 \item School of Civil, Environmental and Chemical Engineering, RMIT University, Melbourne, VIC 3001, Australia,
 \item Max Planck Institute for Dynamics and Self-Organization,37077 G\"ottingen, Germany.
\end{affiliations}

\begin{abstract}
Evaporating liquid droplets are omnipresent in nature and technology, such as in inkjet printing, coating, deposition of materials, medical diagnostics, agriculture, food industry, cosmetics, or spills of  liquids. While the evaporation of pure liquids, liquids with dispersed particles, or even liquid mixtures has intensively been studied over the last two decades, the evaporation of ternary mixtures of liquids with different volatilities and mutual solubilities has not yet been explored. Here we show that the evaporation of such ternary mixtures can trigger a phase transition and the nucleation of microdroplets of one of the components of the mixture. As model system we pick a sessile Ouzo droplet (as known from daily life - a transparent mixture of water, ethanol, and anise oil) and reveal and theoretically explain its four life phases: In phase I, the spherical cap-shaped droplet remains transparent, while the more volatile ethanol is evaporating, preferentially at the rim of the drop due to the singularity there. This leads to a local ethanol concentration reduction and correspondingly to oil droplet nucleation there. This is the beginning of phase II, in which oil microdroplets quickly nucleate in the whole drop, leading to its milky color which typifies the so-called 'Ouzo-effect'. Once all ethanol has evaporated, the drop, which now has a characteristic non-spherical-cap shape, has become clear again, with a water drop sitting on an oil-ring (phase III), finalizing the phase inversion. Finally, in phase IV, also all water has evaporated, leaving behind a tiny spherical cap-shaped oil drop.
\end{abstract}

A coffee drop evaporating on a surface leaves behind 
a roundish stain \cite{deegan1997}. The reason lies in the pinning of the drop on the surface,
together with the singularity of the evaporation rate at the edge of the drop, towards where the colloidal particles of 
the drop are thus transported. This  so-called 'coffee-stain-effect' 
has become paradigmatic for a whole class of problems, and nearly 20 years after Deegan et al.\ \cite{deegan1997}
 presented it to the scientific community, still various questions are open and the problem and its variations keep inspiring the community
\cite{picknett1977,deegan1997,hu2002,sefiane2003,popov2005,cazabat2010,sbonn2006,ristenpart2007,lim2008,ming2008,sefiane2008,liu2008binarydrop,schoenfeld2008,christy2011,gelderblom2011,marin2011,brutin2011,ledesma2014}, 
 %\cite{cazabat2010,marin2011}. 

What happens when an Ouzo drop is evaporating? 
The Greek drink Ouzo (or the French Pastis or the Turkish Raki) consists of an optically 
 transparent ternary mixture of water, ethanol, and anise oil. When served, water is  often added, leading to the nucleation of many tiny oil droplets, which give the drink its milky appearance. This is the so-called Ouzo-effect \cite{vitale2003}. 
As we will see in this paper, also this problem can become paradigmatic, due to its
extremely rich behavior, now for the evaporation-triggered phase separation of ternary liquids and droplet nucleation therein.
%Such ternary liquids have  technological relevance in inkjet printing and 3D manufacturing,  making certain 
%substances such as those with high molecular weight  jettable. 
%This includes the inkjet printing of disposable medical devices, rapid manufacturing
 %applications e.g.\  
  %for polymer LED-based lighting devices, for OLED devices and
  %for solar cells \cite{sirringhaus2000,gans2004,williams2006,dijksman2007,kateri2003}.
  %It also includes the jetting of  ternary  liquids used in medical diagnostics 
  %and biotechnological applications or in the food
 % and cosmetics industry and for coating applications \cite{Hughes2001,Creran2014,Murphy2014,daLuz2015,Yamada2015}. 

The reason for the Ouzo effect lies in the varying solubility of 
  oil in ethanol-water mixtures: With increasing water concentration during the solvent exchange (i.e., water being added), 
 the oil solubility decreases, leading to droplet nucleation in the bulk
 and -- if present -- also on hydrophobic surfaces (so-called surface nanodroplets)
 \cite{zhang2007nanodroplet,lohse2015rmp}.

\section*{Experiments and numerical modelling}
\subsection{Series of events during  evaporation of a sessile Ouzo droplet and their interpretation}
\begin{figure*}[h]
\centering
\includegraphics[width=\textwidth]{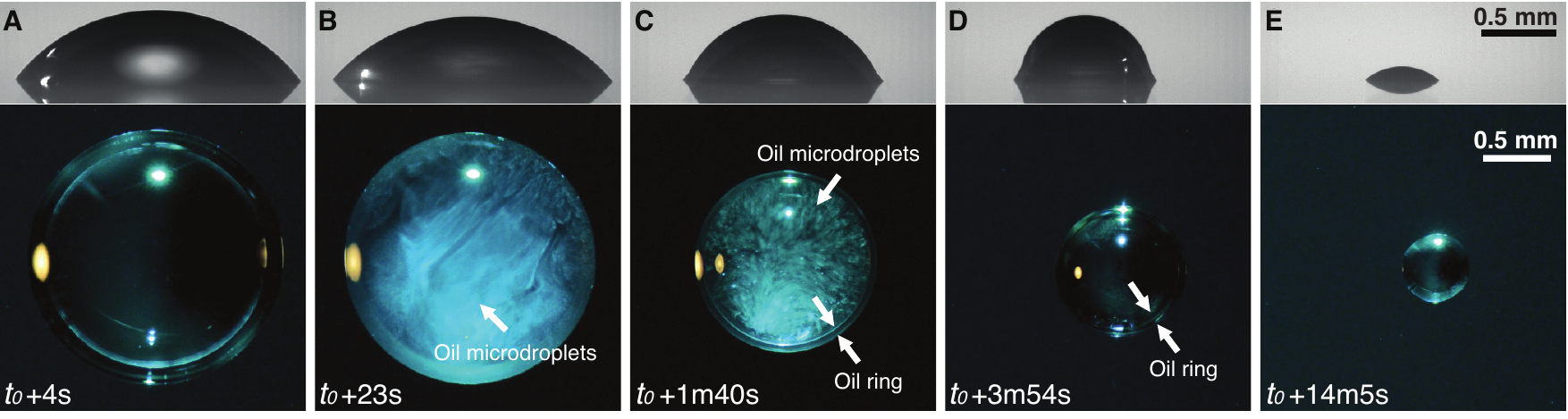}
\caption{Experimental snapshots during the evaporation of an `Ouzo' drop on a flat surface. The initial volume of the drop is \SI{0.7}{\micro\liter} with an initial composition of  \SI{37.24}{\percent} water, \SI{61.06}{\percent} ethanol and \SI{1.70}{\percent} 
anise oil 
(a mixture we refer to as 'Ouzo') in terms of 
weight fractions. 
The time $t_0$ is defined as the moment the needle was pulled out of the drop. 
A time series of the evaporation process can be seen in Videos S1 and S2.
\textbf{(A)} At early times, the Ouzo drop is transparent and has a spherical-cap shape. 
The light ring and spots in the top view image are caused by reflection and refraction of the light source. 
\textbf{(B)} A color transition arises as a result of the Ouzo effect, i.e.\ the nucleation of nano- to micro-sized oil droplets, which are convected by the flow inside the Ouzo drop. The scattering of light at the nucleated 
microdroplets  leads to the  milky coloring of the drop.
\textbf{(C)} The Ouzo drop loses its spherical cap shape due to the appearance of an oil ring. 
The complex transitions from (A) to (C) happen within two and a half minutes, a short time compared to the whole process.
\textbf{(D)} The Ouzo drop is transparent again. Oil microdroplets in the bulk grow big enough to 
sit  on the surface or directly merge 
with the oil ring by convection. 
\textbf{(E)} After around 14 minutes of evaporation, only anise oil is left, now in a spherical cap shape again.
}
\label{fig:snapshots}
\end{figure*}

\begin{figure*}[h]
\centering
\includegraphics[width=\textwidth]{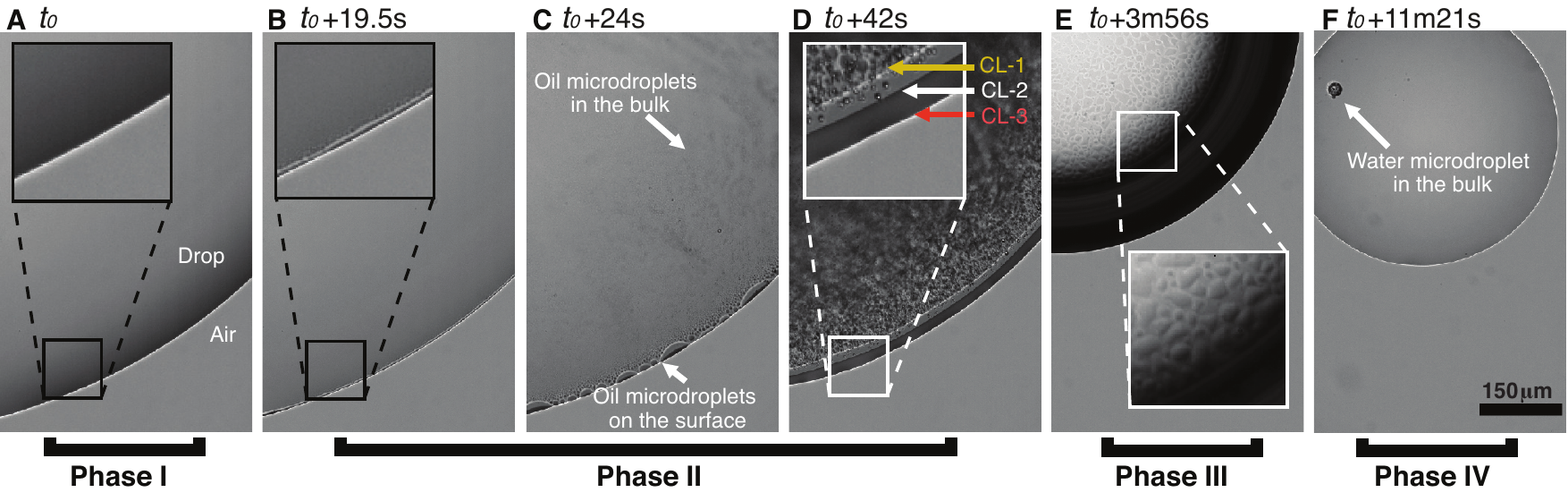}
\caption{Bottom-view snapshots of the contact region of an
 evaporating {\SI{0.7}{\micro\liter}}
  Ouzo drop of the same composition as in Figure \ref{fig:snapshots}.  Again, a video is available as Video S3. 
 %(\SI{32.02}{\percent} water, \SI{66.5}{\percent} ethanol and \SI{1.48}{\percent} anise oil by volume):
 \textbf{(A)}, Phase I: The Ouzo drop is totally transparent with a clearly defined contact line (CL). 
 %Here, $t_0$ represents the time between dropping the drop on the surface and focusing on the contact line.
\textbf{(B)} Phase II: After around \SI{20}{\second}, the contact line is thickened due to the nucleation of oil microdroplets at the rim
as shown in the zoomed-in graph. 
\textbf{(C)} Oil microdroplets nucleated near the contact line are convected throughout the entire drop. 
Meanwhile, the oil microdroplets at the contact line grow and coalesce. 
\textbf{(D)} An oil ring has appeared,
 caused by the deposition of coalesced oil microdroplets on the surface. The zoomed-in graph reveals the presence of three contact lines CL-1, CL-2, and CL-3 near the oil ring, as explained in the main text.  
 The drop is opaque by the presence of numerous oil microdroplets in the bulk. 
\textbf{(E)} Phase III: The outer diameter of the oil ring is smaller, while the thickness is much larger. 
The drop has become  transparent again and many 
merged oil microdroplets on the surface can be observed. 
\textbf{(F)} The drop is transparent with a single contact line CL-3. 
A water microdroplet has been produced as residual of the contracting line CL-2. 
Finally, this remaining water dissolves into the oil and disappears, leaving a homogeneous oil drop (Phase IV).}
\label{fig:bottom}
\end{figure*}

\begin{figure*}[h]
\centering
\includegraphics[width=0.8\textwidth]{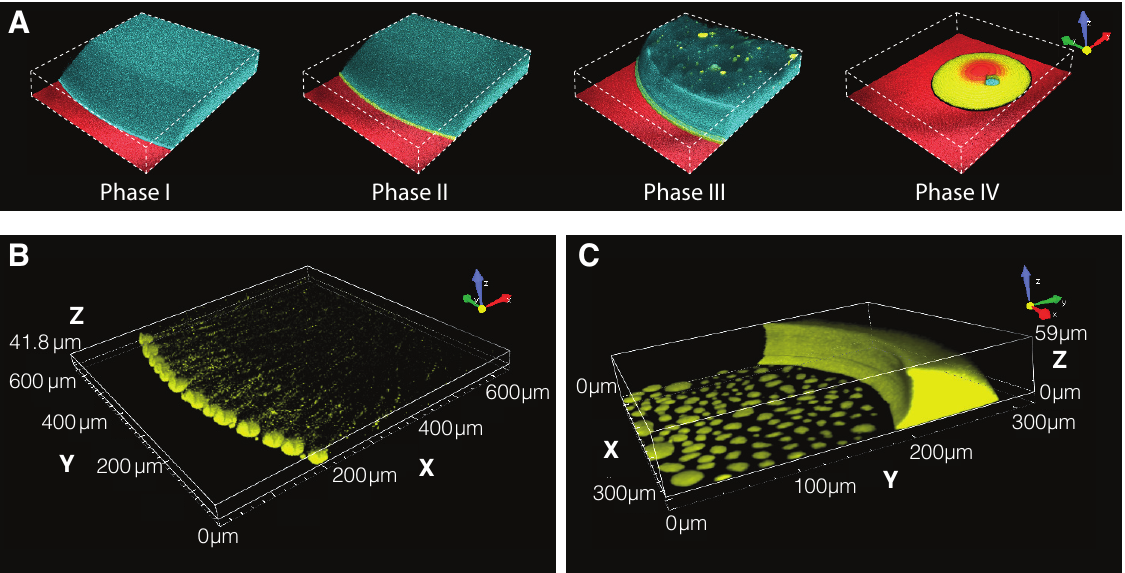}
\caption{Confocal images of the Ouzo drop in different phases. Water/ethanol solution (blue) and oil (yellow) were labeled with different dyes in the confocal experiment.
\textbf{(A)} Morphology of the evaporating Ouzo drop corresponding to four different life phases, taken from a confocal view (Video S4). The scan volume of the confocal microscope is \SI{560}{\micro\meter} \SI{560}{\micro\meter} \SI{90}{\micro\meter}.
\textbf{(B)} The coalesed oil microdroplets on the surface and fresh nucleated oil microdroplets in the bulk were presented in 3D at $t_0 + \SI{26}{\second}$ (early in phase II) .
For the appropriate spatial resolution the 3D images had to be taken over a period of 0.9 s, leading to motion blur of the moving oil microdroplets.
\textbf{(C)} As the oil ring shrinks over time, surface oil microdroplets are destined to be absorbed as shown at $t_0 + \SI{374}{\second}$ (early  in phase III).
A confocal movie of the early nucleation process is  supplied as Video S5.  
 }
\label{fig:conf}
\end{figure*}

\begin{figure*}[h]
\centering
\includegraphics[width=0.7\textwidth]{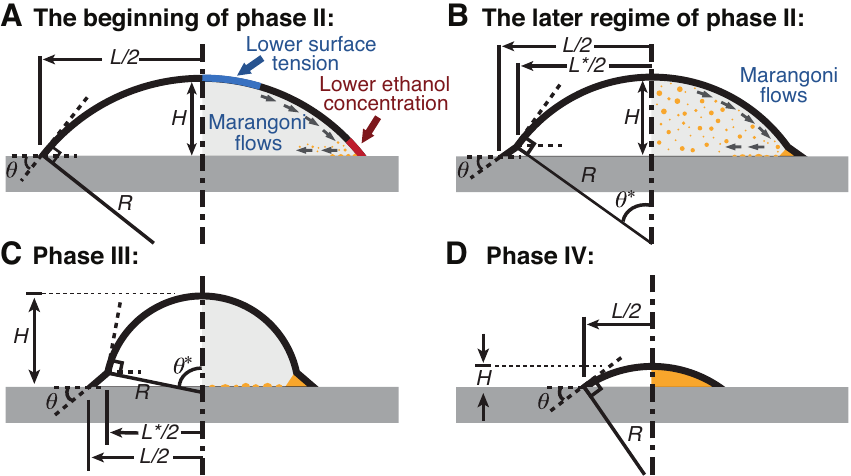}
\caption{Schematics of the Ouzo drop with the definitions of the geometrical parameters at four particular moments.
\textbf{(A)} Due to the preferential evaporation of ethanol near the contact line, the nucleation of oil microdroplets starts in this region. 
The surface tension gradient drives a Marangoni flow that leads to a convection of the oil microdroplets. Despite the non-uniform surface tension, the contour of the drop is well described by a spherical cap with radius $R$.
\textbf{(B)} At later times of regime phase II, the 
oil microdroplets are present in the entire drop and also cover the surface. Meanwhile, 
the oil ring (indicated by the orange triangular region) has appeared, which allows for
the definition of two new geometrical parameters $L^*$ and $\theta^*$. 
\textbf{(C)} After the ethanol content has completely evaporated, the main part of the drop consists of water only. The oil microdroplets in the bulk have coalesced and form a thicker oil ring and larger oil microdroplets on the 
substrate.  Due to the relatively slow evaporation rate of water as compared to ethanol, this stage lasts much longer than  phase II.
\textbf{(D)} Finally, only the non-volatile oil remains after both ethanol and water have evaporated. The sessile 
drop now again has a spherical-cap shape. }
\label{fig:dropsketch}
\end{figure*}

When an Ouzo drop is evaporating, this Ouzo effect is {\it locally}
  triggered by the preferred evaporation of the more volatile ethanol 
 as compared to the less volatile water and the even less volatile oil. As the evaporation rate is highest at the rim of the drop \cite{cazabat2010},
 we expect the oil microdroplets to nucleate there first. Indeed, this is what we see in our experiments, 
 in which we have deposited a $\mu L$ Ouzo drop on a transparent hydrophobic octadecyltrichlorosilane (OTS)-glass surface,
  monitoring its evaporation under ambient conditions with optical imaging synchronized from the top and side 
 (Fig.\  \ref{fig:snapshots} and Videos S1 and S2, experimental setup sketch see Fig. S1%\ref{fig:setupsketch}
 ), from the bottom
 (Fig.\ \ref{fig:bottom} and Video S3)  and confocally (Fig.\ \ref{fig:conf} and Videos S4 and S5). 
 For an illustration of the evaporation process see Figure \ref{fig:dropsketch}.
 At early times, the Ouzo drop is transparent and has a spherical cap shape (Fig.\ \ref{fig:snapshots}A). This is phase I of the evaporation
 process. After about 20 s, indeed microdroplets nucleate at the rim of the drop, as seen in 
 Figure\ \ref{fig:bottom}B or Figure\ \ref{fig:conf}B. This signals the onset of
 phase II, sketched in Figure\ \ref{fig:dropsketch}A: 
  The microdroplets are convected throughout the whole Ouzo drop, giving it its 'milky' 
appearance (Fig.\ \ref{fig:snapshots}B). Due to the
declining ethanol concentration, the liquid becomes oil-oversaturated (cf. Materials and Methods section and Fig. S2).
%\ref{fig:ternarygraph}
 This  oil-oversaturation leads to further 
  oil droplet growth \cite{zhang2015} and coalescence (Fig.\ \ref{fig:bottom}C). 
Finally, an oil ring appears,  caused by the deposition of coalesced oil microdroplets on the surface
(sketch in Fig.\ \ref{fig:dropsketch}B and 
Figs.\ \ref{fig:snapshots}C, 
\ref{fig:bottom}D and \ref{fig:conf}A). 
The zoomed-in graph in Figure \ref{fig:bottom}D and Figure \ref{fig:conf}A reveal
 the presence of three contact lines (CL) near the oil ring: CL-1, where mixture, surface and oil meet,
  CL-2, where mixture, oil and air meet, and 
   CL-3,  where oil, substrate and air meet. The drop is still opaque due to the presence of the numerous oil microdroplets in the bulk. 
However, after about four minutes all ethanol has evaporated. In this phase III, most of the oil droplets have coalesced to an oil ring at the rim
of the drop, which now is transparent again (Figs.\ \ref{fig:snapshots}D, 
\ref{fig:bottom}E, and \ref{fig:conf}C  and sketch in Fig.\ \ref{fig:dropsketch}C).
 In this now phase-inverted phase  the drop has a very characteristic
non-spherical cap-shape, with a water drop sitting on an oil ring. Subsequently, 
the water drop evaporates more and more. The last traces of water are seen as water microdroplets in the bulk of the remaining 
spherical-cap shaped sessile oil drop 
(Fig.\ \ref{fig:bottom}F, phase IV), which now again  has  a single contact line. 
After around 14 minutes of evaporation, only a tiny sessile oil droplet is left (with 1/70th of the original
drop volume), now in spherical cap shape again (Fig.\ \ref{fig:snapshots}E
and sketch Fig.\ \ref{fig:dropsketch}D).

The four life phases of the evaporating Ouzo drop are not only seen visually, but also reflect in various
{\it quantitative} measures of the drop geometry, as extracted from the images of Figures 
\ref{fig:snapshots} and  \ref{fig:bottom}, according to the procedure described in Supporting Information and Fig. S3. %\ref{fig:imageanalysis}.
In Figure \ref{fig:result}A-D we show the measured drop volume $V(t)$, 
its contact diameter $L(t)$ and  the diameter $L^*(t)$ of the water drop sitting on the oil ring,
the corresponding contact angles $\theta (t)$ and $\theta^*(t)$, and the radius of curvature $R(t)$ of the drop. 
The four characteristic phases are separated by three black vertical dashed lines:
Phase I, before the Ouzo effect starts, i.e. before the microdroplets are optically observed at the rim of the drop; phase II, before all ethanol in the drop has evaporated, which is determined from a force balance analysis at CL-2 as detailed in Materials and Methods section; phase III, before the water in the drop has evaporated, i.e. before $\theta (t)$ approaches the contact angle of pure anise oil; and phase IV, when the drop consists of oil only.

After approximately \SI{60}{\second}, the oil ring appeared which is indicated in Figure \ref{fig:result} as a green vertical solid line. From that moment, the evolution of the two additional geometrical parameters $L^*$ and $\theta^*$ is shown. 
In phases I and II, $V(t)$ and $L^{(*)}(t)$ decrease very fast, due to the high evaporation rate of ethanol. Once all
ethanol has evaporated, at the transition from phase II to phase III, there is a sharp reduction in the slope of
 $V(t)$,  $L^{(*)}(t)$, and $R(t)$, which in phase-inverted phase III
 decrease  more slowly due to the lower evaporation rate of water.
 In this regime, a force balance holding at CL-2 reaches its steady state (Fig. S4%\ref{fig:def}
 ).
In the final phase, $V(t)$  converges to the initial volume of the anise oil (zoomed-in graph in Fig.\
\ref{fig:result}A) and $\theta(t)$ approaches the contact angle of pure anise oil (Fig.\ \ref{fig:result}C). 

\subsection{Numerical modelling of the evaporation process and its quantitative understanding}
More quantitative insight is gained from numerically modelling the evaporation process of the Ouzo drop (Video S6).
Our numerical model is based on an axisymmetric  lubrication approximation in the spirit of
the evaporating coffee-stain lubrication models of 
refs.\ \cite{deegan1997,deegan2000,popov2005,gelderblom2011}, but now for a 
multi-component liquid. The relative mass fractions are governed by a convection-diffusion equation,
with a sink-term at the air-drop interface, reflecting evaporation, and ethanol-concentration-dependent
material parameters such as density, diffusivity, viscosity, surface tension, and activity coefficients
(quantifying the evaporation rate). These composition-dependent properties are depicted in Fig. S6. %\ref{fig:model:propfits}.
The Ouzo drop is described assuming axial symmetry, 
with the  liquid-air interface given by the height function $h(r,t)$ and 
 the fluid velocity $\vec{v}=(u,w)$ (cf. Fig.~S5 %\ref{fig:model:scheme}
 ). Details of the model are given in Supporting Information.
 
 The fundamental difference between the evaporation of a pure liquid \cite{deegan2000} 
 and that of a mixture is the vapor-liquid equilibrium. While in the case of a pure 
 liquid 
 $\alpha$ the vapor concentration $c_{\alpha}$ (mass per volume) directly above the liquid-air interface is saturated, i.e. $c_{\alpha}=c_{\alpha,\text{sat}}$, it is lower for the case of mixtures. The relation between liquid composition and vapor composition is expressed by Raoult's law.
As in the evaporation model for a pure liquid \cite{deegan2000}, 
the evaporation rate $J_\alpha$ is obtained by solving the quasi-steady vapor-diffusion 
$\nabla^2 c_{\alpha}=0$ in the gas phase with the boundary conditions 
given by Raoult's law above the drop, by the no-flux condition 
$\left.\partial_z c_{\alpha}\right|_{r>L/2,z=0}=0$ at the drop-free substrate, 
and far away from the drop by the given vapor concentrations $c_\alpha = 0$ for ethanol and 
$c_{\alpha}=c_{\alpha,\infty}=RH_{\alpha}c_{\alpha,\text{sat}}$ for water, where
  $RH_{\alpha}$ is the relative humidity. The relative humidity can be measured to some limited precision, but here
  had to be corrected for to better describe the experimental data, as detailed in Materials and Methods section.
 Finally, the evaporation rates are  given by $J_\alpha=-D_{\alpha,\text{air}}\partial_n c_{\alpha}$ with the vapor diffusion coefficients $D_{\alpha,\text{air}}$ of $\alpha$ in air.
In contrast to the evaporation of a pure fluid, the evaporation rate of a mixture component does not only depend on the geometric shape of the drop, but also on the entire composition along the liquid-air interface.
The resulting $r$-dependent height loss due to evaporation is given in Supporting Information. 

\begin{figure*}[h]
\centering
\includegraphics[width=0.7\textwidth]{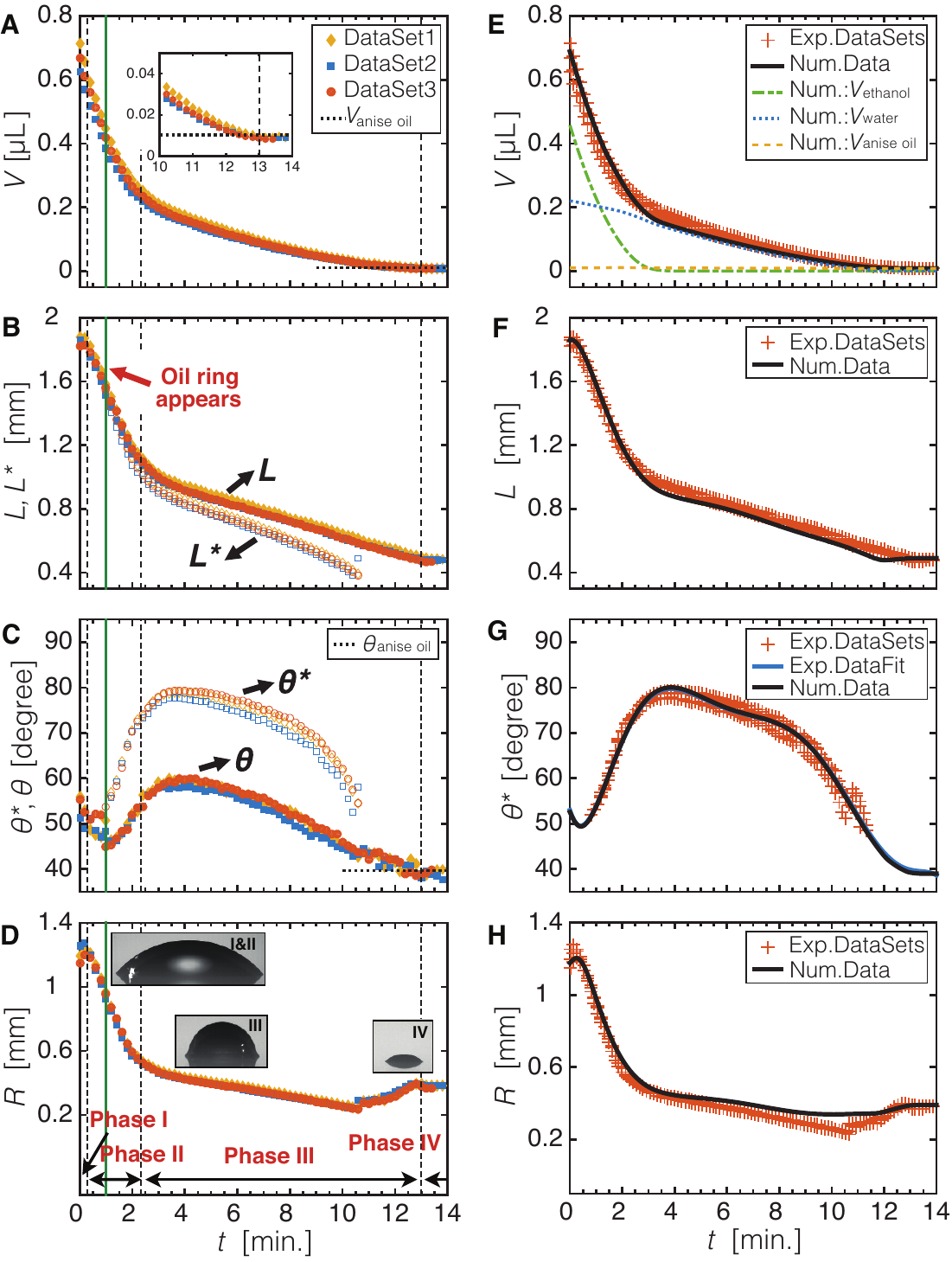}
\caption{Experimental (A-D) and numerical (E-H) results for the temporal evolution of the geometrical parameters: Volume $V$ (\textbf{A}, \textbf{E}), lateral sizes $L$ and $L^*$ (\textbf{B}, \textbf{F}), 
contact angles $\theta$ and $\theta^*$ (\textbf{C}, \textbf{G}), and radius of curvature $R$ (\textbf{D}, \textbf{H}). 
The vertical dashed lines mark the transition from one phase to another. }
\label{fig:result}
\end{figure*}

In the simulations, the fitted experimental data $\theta^*$ (shown in Fig. \ref{fig:result}G) were
 used as the time-dependent contact angle.
The  quantitative measures of the drop geometry resulting from the numerical simulations 
 are shown in 
Figures \ref{fig:result}E, \ref{fig:result}F, and \ref{fig:result}H, together with the experimental data, showing excellent 
quantitative agreement.
From  Figure\ \ref{fig:result}E, which next to the total volume 
$V(t)$ also shows the partial volumes of the three components water, ethanol, and oil, 
we can reconfirm 
 that the volume loss is initially mainly due to the evaporation of ethanol (phase I and II), followed by a slower evaporation of the remaining water (phase III). Finally, only the  tiny non-volatile oil droplet remains (phase IV).

\begin{figure*}[h]
\begin{center}
\includegraphics[width=0.8\textwidth]{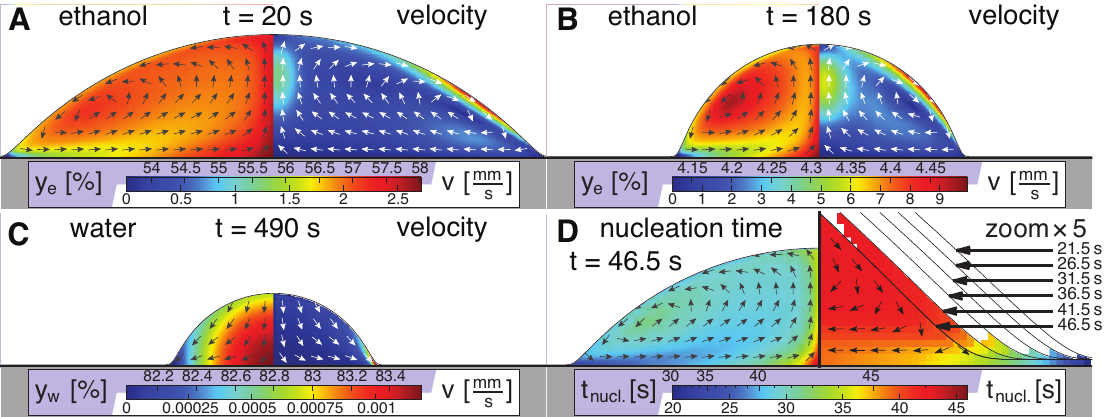}
\caption{Snapshots of the numerical results at three different times 
t = 20 s (A), t = 180 s (B) and t = 490 s (C). \textbf{(A,B)} Mass fraction of ethanol $y_{\text{e}}(r,z,t) $ and 
 fluid velocity field $\vec v(r,z,t) $, whose
  direction is indicated by the arrows and whose modulus by the color-code. 
  At the later time t=490 s in \textbf{(C)},  the water concentration is plotted instead of the ethanol concentration (which then is 
  close to zero), 
  again together with the velocity field. 
   \textbf{(D)} Oil droplet nucleation time $t_{\text{nucl.}}$. 
  The right side shows a zoom-in of the region around the rim. 
  A movie of the numerical simulation is  supplied as Video S6.}
\label{fig:model:dropview}
\end{center}
\end{figure*}

Our  numerical simulations of the process  allow us to deduce the fully spatially resolved 
mass fraction and velocity  fields,  $y_\alpha(r,z, t)$ and 
$\vec v (r,z,t)$, respectively. In Figures\ \ref{fig:model:dropview}a and \ref{fig:model:dropview}b we show the ethanol mass fraction
$y_{\text{e}}(r,z, t)$ and the velocity field
$\vec v (r,z,t)$ for two different times $t = \SI{20}{\second}$ and $t = \SI{180}{\second}$. 
It is clearly visible how the preferential evaporation of ethanol near the contact line, which leads 
to a larger surface tension there, 
drives a fast Marangoni 
flow. As a consequence, ethanol is quickly replenished at the liquid-air interface and can completely evaporate. 
We note that
 the direction of the convection roll inside the drop is opposite to the case of a pure liquid, where 
 the flow goes outwards at the bottom of the drop and inwards at the liquid-gas interface 
 \cite{deegan1997,deegan2000,gelderblom2011}. We also note that 
 the ethanol concentration differences are relatively small -- in the beginning about 3\% and 
 later not more than 0.5\% -- 
 but nonetheless sufficient to drive a strong Marangoni flow with velocities up to the order of 10 mm/s. 
 Due to the high contact angle during phases II and III, the lubrication approximation predicts the precise values of the velocity only to a limited accuracy. The qualitative flow field and the order of magnitude, however, have been validated by a comparison with the corresponding non-approximated Stokes flow at individual time steps. 
 Figure \ref{fig:model:dropview}C shows the 
 water mass fraction $y_{\text{w}}(r,z,t) $ for $t = \SI{46.5}{\second}$, since at these later times 
 ethanol is virtually not present anymore, again together with the velocity field, which is now again
 outwards directly above the substrate. 
 
 Finally, in Figure \ref{fig:model:dropview}D we show the oil droplet nucleation time $t_{\text{nucl.}}$, which is defined as the moment 
  when the local composition crosses the phase separation curve and enters the Ouzo region (see Fig. S2%\ref{fig:ternarygraph}
  A). 
According to the numerical results, the oil droplet nucleation starts at $\SI{20}{\second}$ near the contact line,
in perfect agreement with our experimental findings,
 and nucleation is possible in the entire droplet at $t=\SI{46.5}{\second}$.

\section*{Conclusions and outlook}
In summary, we have experimentally and numerically
studied  the evaporation of a millimetre sized sessile Ouzo drop  
on a hydrophobic substrate. How stimulating it can be to study the evaporation of alcoholic drinks has interestingly  also  been shown 
in a very recent parallel but independent work by Kim et al.\ \cite{Kim2016a}, who studied  the drying  of 
{\it whisky} droplets, which give a uniform deposition pattern. For that system suspended material and 
surface-absorbed macromoleculars play a major role and offer a physicochemical avenue for the control of coatings.
From our point of view,
just as the evaporating whisky droplet, also the  evaporating ouzo droplet can advance our 
scientific understanding
of complex flow phenomena and phase transitions and their interaction. 
%, revealing its four life phases. 
In this paper we have observed  evaporation-triggered phase transitions and the nucleation of oil microdroplets, first 
at the edge of the Ouzo drop and then allover, followed by a phase inversion, and altogether four different life phases of the ouzo drop, which 
serves as paradigmatic model system for 
ternary mixtures of liquids with different volatilities and mutual solubilities. Here, water as the second but most volatile 
liquid (after the very quickly  evaporating ethanol) also evaporates in about ten  minutes, leaving behind a tiny drop
of anise oil. % , which has only 1/70th of the original drop volume.
 For other ternary mixtures only one liquid may be volatile, implying phase III with a binary mixture and nucleated
microdroplets of one liquid and its peculiar optical properties would be the final state. 

Tuning and optimizing the 
material and chemical 
properties  of the individual liquids in the ternary mixture such as volatilities and mutual solubilities 
and polymerisibility (e.g.\ under UV exposure such as in ref.\ \cite{zhang2012softmatter}) offers
a plethora of applications for medical diagnostics, the controlled deposition of complex liquids in the food and cosmetic industry and 
for coating applications \cite{Hughes2001,Creran2014,Murphy2014,daLuz2015,Yamada2015}, in agriculture,
or the food or cosmetics industry,  
for inkjet printing of LED or OLED devices and   solar cells  \cite{sirringhaus2000,gans2004,williams2006,dijksman2007,kateri2003}, and for rapid manufacturing. 
Here we studied the deposition on smooth surfaces, but pre-patterning the surface with hydrophobic patches
\cite{bao2015} offers even further opportunities, by directing the nucleation of nano- or  microdroplets at will, allowing for
the self-organised {\it bottom-up} construction of structures.

\section*{Material}
\subsection{Ternary diagram and initial composition of the Ouzo drop} 
The ternary liquid of the Ouzo drop in this study was the mixture of Milli-Q water (produced by a Reference A+ system, Merck Millipore, at \SI{18.2}{\mega\ohm\centi\meter}), ethanol (EMD Millipore, Ethanol absolute for analysis) and anise oil (Aldrich, Anise oil).
The ternary diagram of the mixture was titrated at a temperature of \SI{22}{\celsius}, which is similar to the environmental temperature during the evaporation experiment. 
\SI{21} groups of ethanol and anise oil mixtures with different component weight ratios were properly prepared to be used as titrants (see Table S1).
The volume of water (titrate) was precisely measured by a motorised syringe pump (Harvard, PHD 2000). For each ethanol and anise oil mixture, a phase-separation point was determined as shown in Fig. S2A. Photographs of the macrosuspensions corresponding to the different phase-separation points were taken. Thereby, the stability of the macrosuspension along the phase separation curve was determined (Fig. S2B). Starting with point g, the homogeneous macrosuspension is not stable anymore.
The part of the curve with a stable macrosuspension was identified as the boundary of the Ouzo region in the ternary diagram, which is labeled Ouzo range. 
According to the ternary diagram, the initial composition of the Ouzo drop was chosen as \SI{37.24}{\percent} water, \SI{61.06}{\percent} ethanol and \SI{1.70}{\percent} anise oil in terms of weight fractions, which is indicated by the black star in Fig. S2A.
Starting from this initial point, the drop composition is guaranteed to cross the phase separation curve and enter the Ouzo region during the evaporation process. A black dotted line in the magnified subfigure of Fig. S2A shows the numerically obtained temporal evolution of the composition near the contact line of the Ouzo drop. 

\subsection{Experimental methods} 
A \SI{0.7}{\micro\liter} Ouzo drop (\SI{37.24}{\percent} water, \SI{61.06}{\percent} ethanol and \SI{1.70}{\percent} anise oil in terms of weight fractions) was produced through a custom needle (Hamilton, O.D.$\times$I.D. (mm): 0.21$\times$0.11) by a motorised syringe pump (Harvard, PHD 2000). 
The whole evolution of the Ouzo drop was observed by two synchronised cameras, one (Photron Fastcam SA-X2 64GB, 50 fps at 1,024 $\times$ 1,024 pixel resolution) affixed with a high-magnification zoom lens system (Thorlabs, MVL12X3Z) for side-view recordings and another (Nikon D800E,  25fps at 1,920 $\times$ 1,080 pixel resolution) affixed with an identical lens system for top-view recordings (Figure 1, Videos S1 and S2).
The temperature around the evaporating drop was measured using a thermometer sensor. The relative humidity in the lab was measured with a standard hygrometer ($\pm$\SI{3}{\percent}RH for \SI{35}{\percent}$\sim$\SI{70}{\percent}RH at \SI{20}{\celsius})
The temperature of the three experimental datasets in Figure 5 was between \SI{21}{\celsius} and \SI{22.5}{\celsius}. The relative humidity was around \SI{40}{\percent}. 
The image analysis was performed by custom-made MATLAB codes.
In order to have a detailed observation of the evolutionary process at the rim of the Ouzo drop, an inverted microscope (Olympus GX51) was used to focus on the contact region. A fast speed camera (Photron Fastcam SA-X2 64GB, 50 fps at 1,024 $\times$ 1,024 pixel resolution) was connected to the microscope with an intermediate tube. Figure 2 and Video S3 were taken with a $20\times$ long working m-plan fluorite objective (Olympus MPLFLN20XBD, Wd = 3.0 mm, NA = 0.45).
Besides 2D imaging, we also took advantage of a confocal microscope (Nikon Confocal Microscopes A1 system) in stereo-imaging. A real-time observation was carried out to monitor the movement of the oil droplets due to the convective flow and the formation of oil ring in a 3D view. A $20\times$ air objective (CFI Plan Apochromat VC $20\times$/0.75 DIC, NA = 0.75, WD = 1.0 mm) and a $40\times$ air objective (CFI Plan Fluor $40\times$/0.75 DIC, NA = 0.75, WD = 0.66 mm) were employed to take Figures 3A, 3B and Figure 3C, respectively. In Figures 3B and 3C and Video S5, anise oil was labeled by Nile Red (Microscopy grade, Sigma-Aldrich, Netherlands). 
In Figure3A and Video S4, in order to simultaneously label oil and solution with different color dyes during the whole evaporating process, anise oil was replaced by trans-Anethole oil (\SI{99}{\percent}, Sigma-Aldrich, Netherlands) labeled by perylene (sublimed grade, $\geq$\SI{99.5}{\percent}, Sigma-Aldrich, Netherlands) in yellow color. Water/ethanol mixture was labeled by fluorescein 5(6)-isothiocyanate (High performance liquid chromatography, Sigma-Aldrich, Netherlands) in blue color.

\subsection{Definitions of the four life phases of an evaporating Ouzo drop}
We divided the Ouzo drop evaporation process into four phases: Phase I is defined as the initial regime, before the critical phase separation composition is attained at the contact line. Phase II is the time from the initial occurrence of the oil nucleation until the complete evaporation of the ethanol component. Phase III is the regime when the remaining water amount in the drop evaporates. The final phase IV is the period after the the remaining water has evaporated. The first black vertical dashed lines (separation between phases I and II) and the third one (separation of phases III and IV) in Figure 5 were able to be optically determined from the top or bottom view video recordings. However, the transition between phase II and phase III cannot be detected from the video recordings. 
Instead, the second black vertical dashed line in Figure 5 was determined from an equilibrium analysis as a simplified model (cf. Fig. S4A): at the air-mixture-oil contact line (CL-2 in Figs. 2B, 3A and 3C), a force balance holds. The influence of the line tension on the balance can be neglected \cite{torza1970three}. Each variation of the composition in the drop alters the equilibrium of this balance\cite{bazhlekov1997numerical,torza1970three}. At the moment when ethanol has completely evaporated, this equilibrium attains its steady state. From that moment, the three phases which meet at the contact line CL-2 are water from the liquid of the drop, anise oil from the oil ring and air from the surroundings. The composition of the air phase near the contact line CL-2 is assumed to be constant. Hence, the angle between the mixture-air interface and the oil-air interface has to be constant. Mathematically speaking, this means that $\Delta\theta$ has to be a constant. The quantity $\Delta\theta$ was estimated by the subtraction $\theta^*-\theta$, since the dimension of the oil-air interface is small in the initial part of phase III. In Fig. S4B, the evolution of $\Delta\theta$ as a function of time is shown. It is clearly visible that after a rapid increase $\Delta\theta$ remains constant for a very long time. Therefore, we fitted $\Delta\theta$ from time $t_a$ to time $t_z=\SI{480}{\second}$ by a constant $c$. The inserted graph in fig. S4B shows the relation between $c(t_a)$ and $t_a$. We selected the time $t_a=\SI{140}{\second}$ as the separation moment between phase II and phase III.

\subsection{Numerical model} 
The evolution of the drop shape $h(r,t)$ (cf. Fig. S5) is solved by a diagonally-implicit Runge-Kutta method, the vapor diffusion-limited evaporation rates are calculated by a boundary element method and the convection-diffusion equations for the composition are treated with an upwind finite differences scheme. For the composition-dependency of the mass density, the surface tension, the diffusion coefficient, the viscosity and the activity coefficients, we fitted experimental data of water-ethanol mixtures or used appropriate models (cf. Fig. S6). Details can be found in the Supporting Information. Our model was validated for the case of pure water by comparison with the experimental data of Gelderblom et al. \cite{gelderblom2011}.

\subsection{Determination of the relative humidity}For the numerical simulation, we have assumed a temperature of $T=\SI{21}{\celsius}$ and a relative humidity of $RH_{\text{e}}=0$ for ethanol. Since the experimental determination of the relative humidity $RH_{\text{w}}$ of water is error-prone, we have determined it as follows: At the beginning of phase III, the drop consists almost entirely of water, since the ethanol content has already evaporated and the amount of oil is still small in comparison to the remaining water volume. Therefore, we used our numerical model to fit $RH_{\text{w}}$ based on the experimental data for the volume evolution $V(t)$ during the time from $t=\SI{140}{\second}$ to $t=\SI{300}{\second}$. The resulting water humidity reads $RH_{\text{w}}=\SI{63}{\percent}$.

\begin{addendum}
 \item We thank Michel Versluis for invaluable advice on imaging and Shuhua Peng for preparing the substrates. 
D.L.\ gratefully  acknowledges financial support through an
ERC Advanced Grant and the NWO-Spinoza programme.  
H.T. thanks for the financial support from china scholorship council (CSC, file No.201406890017).
H.K. and C.D. gratefully acknowledge financial support by the Dutch Technology Foundation STW.
 \item[Competing Interests] The authors declare that they have no
competing financial interests.
 \item[Correspondence Email]  d.lohse@utwente.nl or xuehua.zhang@rmit.edu.au.
\end{addendum}

\noindent{\bf
Author Contributions}
X.Z. and D.L. conceived the project. H.T., X.Z. and D.L. planned the experiments. P.L. and H.T. performed confocal microscopy experiments. H.T. performed all other experiments in the study. H.T. and D.L. analyzed the experimental data. H.K. and C.D. developed and implemented the numerical model and performed the corresponding simulations. D.L., C.D. and H.T. wrote the paper.
All authors discussed the results and commented on the manuscript.
\\

\clearpage
\newpage
\section*{Supporting Information}
\subsection{1. Hydrophobic octadecyltrichlorosilane (OTS)-glass surface}
The glass substrate the drop was placed on is hydrophobic, being coated with an octadecyltrichlorosilane monolayer (OTS) (made at RMIT, Australia). The advancing and receding contact angle of water on this substrate are \SI{112}{\degree} and \SI{98}{\degree}, respectively. The contact angle of the anise oil used in the experiments on this substrate is \SI{39.6}{\degree}, measured with a video-based optical contact angle measuring system (DataPhysics OCA15 Pro). Before being used, the substrates were cleaned by 15-mins sonication in \SI{99.8}{\percent} ethanol and 5-mins in Milli-Q water sequentially, and subsequently dried with compressed N$_2$ flow for 2 minutes before each experiment. 

\subsection{2. Image analysis and data calculation}
The image analysis was performed by custom-made MATLAB codes, through which all the geometric parameters at every frame were successfully determined, such as drop volume $V$, contact angles $\theta$ and $\theta^*$, lateral sizes $L$ and $L^*$ and droplet height $H$ (cf. Fig. S3A).
The drop volume was calculated by adding the volumes of horizontal disk layers, assuming rotational symmetry of each layer with respect to the vertical axis.
The contact angle $\theta$, between the blue and green lines in Fig. S3A, was estimated from the profile at the contact region by polynomial fits, while $\theta^*$, between the red and yellow lines, was calculated by a spherical cap approximation (purple circle). 
The drop contour above the oil ring was also fitted by elliptical fits. Since the drop size is smaller than the capillary length $\kappa^{-1}=\sqrt{\frac{\gamma}{\rho g}}$ (\SI{2.7}{\milli\meter} for water, \SI{1.7}{\milli\meter} for ethanol), the ellipticity of the top cap, defined as the ratio between the difference of the two semi-axes and the radius, was always below \SI{10}{\percent} during phase II (Fig. S3B). After around 11 minutes, both the spherical cap approximation and the elliptical fittings for the water contour above the oil ring were not sufficiently accurate. The water drop diameter $L^*$ was too small (less than \SI{0.4}{\milli\meter}) and there were not enough pixels to perform the contour fits. Therefore, we stopped calculating $\theta^*$ from a spherical cap approximation at around 11 minutes, when the ellipticity exceeds \SI{10}{\percent}.

\subsection{3. Variation of oil ring contact line}
Remarkably, $\theta$ is not constant during phase III as shown in Figure 5C in the main part. This is caused by the water saturation variation at the air-oil-substrate contact line. The water diffusion speed from the mixture to the oil ring and the speed from the oil ring to the air co-determined the water saturation in the oil ring. Our other experimental work, at a higher ambient relative humidity, which attenuates the water diffusion from oil ring to air, has shown that $\theta$ is constant for a long time in phase III after the same initial increase in phase II. Since a detailed discussion of these experimental results is beyond the scope of this paper, they are not shown here.

\subsection{4. Artificial light signal in confocal images}
In Figure 3B of the main part and Video S5, the vertical blobs above surface oil microdroplets are artificial signals caused by light reflection. At early phase II, the oil microdroplets in the bulk and on the surface do not have enough dye and require a strong laser intensity  to be visualised. The oil-air interfaces on the surface oil microdrolets act as a mirror reflecting the real light signals in the microdroplets. In addition, the artificial light signals caused by reflection are even more intense than the real light signals emitted from the oil microdroplets in the bulk. Hence, these artifacts cannot be suppressed by an appropriate adjustment of the brightness, the contrast or the gamma correction.

In the scanning of Figure 3C, each 2D image of the Z-stack was averaged by 4 images in order to reduce the noise and detect the surface oil microdroplets sharper, but at the expenses of scanning time. Even though the evaporating process in phase III is relatively slow, the \SI{16}{\second} scanning time for one 3D image still leads to a stripe-like distortion of the oil ring.

\subsection{5. Numerical model}
Our numerical model is based on an axisymmetric multi-component lubrication approximation. The Ouzo drop is described in cylinder coordinates $(r,z)$ with the fluid velocity $\mathbf{v}=(u,w)$, where the liquid-air interface is given by the height function $h(r,t)$. A schematic illustration is depicted in Fig.~S5.
%In the following, we focus on the liquid phase only, i.e. $0<z<h(r,t)$, whereas the gas phase becomes relevant again for the evaporation model in section \ref{sec:model:evaprate}.

\subsection{5.1. Liquid composition and local physical properties}
The liquid composition is denoted in terms of mass fractions $y_{\alpha}(r,z,t)$ with $y_{\text{w}}+y_{\text{e}}+y_{\text{a}}=1$. Here, $\alpha=\text{w},\text{e},\text{a}$ stands for the species water, ethanol and anise oil, respectively. 

During the initial phase of the evaporation, in particular when the oil nucleation has not set in yet, the presence of anise oil can be neglected ($y_{\text{a}}(t{=}0)=0.017$). The physical properties of the liquid, i.e. the mass density $\rho$, the surface tension $\sigma$, the dynamic viscosity $\mu$ and the mutual diffusion coefficient $D$, are therefore assumed to be given by those from binary water-ethanol mixtures. Since the nucleated oil droplets are initially small compared to the entire drop size, this assumption will also hold true at intermediate times. We have fitted experimental data of water-ethanol mixtures to incorporate the compositional dependence of $\rho$, $\sigma$, $\mu$ and $D$ into our model (cf. Fig.~S6).

The mass fractions are governed by the following convection-diffusion equation, where the diffusive fluxes in the ternary mixture are assumed to be in the Fickian limit with the diffusion coefficient $D$ of a binary water-ethanol mixture:

\begin{equation}
\rho\left(\partial_t y_{\alpha} +\mathbf{v}\cdot \nabla y_{\alpha}\right)= \nabla\cdot \left(\rho D\nabla y_{\alpha}\right)+J_{\alpha}\delta_{\Gamma}\,.
\label{eq:model:massfraccde}
\end{equation}
The mass flux source term $J_{\alpha}$ is only present at the liquid-air interface (with the interface delta function $\delta_{\Gamma}$) and stems from the evaporation of species $\alpha$ (cf. section 5.3).

%%%%
\subsection{5.2. Lubrication approximation}
Due to the different mass densities of the liquid constituents, the flow in the drop is subject to the full compressible Navier-Stokes momentum equation along with the mass conservations for the individual species. An energy equation is not considered, i.e. the drop is assumed to be isothermal at room temperature $T$, since the dominant mechanism for the Marangoni flow in the drop is the strong dependence of the surface tension $\sigma(r,t)$ on the local liquid composition.

Due to the size of the drop, the influence of gravity can be neglected. The pressure is therefore constituted by the Laplace pressure which results, at least for a homogeneous surface tension, in a spherical cap equilibrium shape:
\begin{equation}
p_\text{L}(r,t)=-\sigma(r,t)\,\frac{1}{r}\,\partial_r \left(\frac{r\,\partial_rh(r,t)}{\sqrt{1+\left(\partial_rh(r,t)\right)^2}}\right)\,.
\label{eq:model:laplpressure}
\end{equation}
In the spirit of large eddy simulations, we introduce a numerical cut-off towards the microscopic scales that are relevant near the free moving contact line. To that end, a precursor film with thickness $h^*=L(t=0)/100$ and a corresponding disjoining pressure 
\begin{equation}
\Pi(r,t)=-\frac{\sigma(r,t)\,\theta_{\text{e}}^2}{2h^*}\frac{(n-1)(m-1)}{(n-m)} \left( \left(\frac{h^*}{h}\right)^n - \left(\frac{h^*}{h}\right)^m \right)
\label{eq:model:disjpressure}
\end{equation}
with $n=3$ and $m=2$ are taken into account, i.e. $p=p_\text{L}+\Pi$ \cite{Schwartz1998a}. Here, $\theta_{\text{e}}$ is the equilibrium contact angle. However, since the exact physical interaction of the liquid-air interface with the deposited oil ring at the contact line is not known in detail, we have fitted the experimental data for $\theta^*(t)$ and adjust the parameter $\theta_{\text{e}}(t)$ according to this fit in such way that the contact angle resulting from the numerical model resembles the experimental data.

Applying lubrication theory on the momentum equation yields the following set of governing equations:
\begin{align}
\partial_r p&=\partial_z \left(\mu \partial_z u \right)\label{eq:model:lubric}  \\
\partial_z p&=0 \,, \label{eq:model:lubricpress} \\
\partial_t \rho+\frac{1}{r}\partial_r  \left(r\rho u \right)+\partial_z  \left(\rho w\right) &=0 \,.
\label{eq:model:lubricconti}
\end{align}

An inhomogeneous composition along the liquid-air interface causes a shear stress, which reads in the order of the lubrication theory $\mu\partial_zu=\partial_r\sigma$.
With the no-slip condition at the substrate $z=0$, the radial velocity is given by
\begin{equation}
u(r,z,t)=\int^z_0  \frac{(-\partial_rp(r,t))\left(h(r,t)-z'\right)+\partial_r \sigma(r,t)}{\mu(r,z',t)}\: \mathrm{d}z' \,.
\label{eq:model:velor}
\end{equation}
The axial velocity $w$ is obtained from \eqref{eq:model:lubricconti} and, consequently, the drop shape evolves according to
\begin{equation}
\partial_t h(r,t)=\frac{1}{\left.\rho\right|_{z=h(r,t)}}\left[-\frac{1}{r}\partial_r  \int^{h(r,t)}_0 r\rho (r,z,t)u(r,z,t) \:\mathrm{d}z
-\int_0^{h(r,t)}\partial_t \rho(r,z,t) \:\mathrm{d}z \right]+ w_{\text{evap}}(r,t)\,,
\label{eq:model:hevo}
\end{equation}
where the height loss $w_{\text{evap}}$ stems from evaporation (cf. section 5.3). 

The present model cannot account directly for the deposited oil ring and possible interactions of the nucleated oil droplets with the flow are also not taken into account. However, in the initial regime, these aspects will be not relevant.

%%%%
\subsection{5.3. Evaporation model}
\label{sec:model:evaprate}
While the evaporation of anise oil can be neglected, the well-established vapor-diffusion limited evaporation model for pure fluids of Deegan et al.\cite{deegan1997,deegan2000} and Popov \cite{popov2005} has to be generalized to mixtures. The fundamental difference between a pure liquid and a mixture is the vapor-liquid equilibrium. While in the case of a pure fluid $\alpha$ the vapor concentration $c_{\alpha}$ (mass per volume) directly above the liquid-air interface is saturated, i.e. $c_{\alpha}=c_{\alpha,\text{sat}}$, it is lower for the case of mixtures. The relation between liquid composition and vapor composition is expressed by Raoult's law, which can be written by the use of the ideal gas law as the boundary condition
\begin{equation}
c_{\alpha}(r,t)=\gamma_{\alpha}(r,t)x_{\alpha}(r,t) c_{\alpha,\text{sat}}\qquad\text{at}\qquad z=h(r,t),\,r<L/2\,.
\label{eq:model:raoults}
\end{equation}
Here, $x_{\alpha}$ is the mole fraction of component $\alpha$ in the liquid and $\gamma_{\alpha}$ is the activity coefficient which comprises possible non-idealities of the mixture. 
As in the evaporation model for a pure fluid, the evaporation rate $J_\alpha$ is obtained by solving the quasi-steady vapor-diffusion $\nabla^2 c_{\alpha}=0$ in the gas phase with the boundary conditions \eqref{eq:model:raoults}, $\left.\partial_z c_{\alpha}\right|_{r>L/2,z=0}=0$ and $c_{\alpha}=c_{\alpha,\infty}=RH_{\alpha}c_{\alpha,\text{sat}}$ far away from the drop (with relative humidity $RH_{\alpha}$). Finally, the evaporation rate is given by $J_\alpha=-D_{\alpha,\text{air}}\partial_n c_{\alpha}$ with the vapor diffusion coefficient $D_{\alpha,\text{air}}$ of $\alpha$ in air.
In contrast to the evaporation of a pure fluid, the evaporation rate of a mixture component does not only depend on the geometric shape of the drop, but also on the entire composition along the liquid-air interface.
The height loss velocity in \eqref{eq:model:hevo} is finally given by
\begin{equation}
w_{\text{evap}}(r,t)=-\frac{J_\text{w}(r,t)+J_\text{e}(r,t)}{\left.\rho\right|_{z=h}(r,t)}\sqrt{1+\left(\partial_r h(r,t)\right)^2}\,.
\label{eq:model:velov}
\end{equation}

The values for $c_{\text{w},\text{sat}}$ and $c_{\text{e},\text{sat}}$ for water and ethanol vapor, respectively, were calculated based on the ideal gas law and the Antoine equation, whereas the vapor diffusivities read $D_{\text{w},\text{air}}=\SI{0.246}{\centi\meter^2\per\second}$ \cite{gelderblom2011} and $D_{\text{e},\text{air}}=\SI{0.135}{\centi\meter^2\per\second}$ \cite{Lee1954a}, respectively . The activity coefficients are depicted in Fig.~S6.

\setcounter{figure}{0}
\makeatletter\renewcommand{\figurename}{Fig. S}\makeatother

\clearpage
\begin{figure*}[h]
\centering
\includegraphics[width=0.9\textwidth]{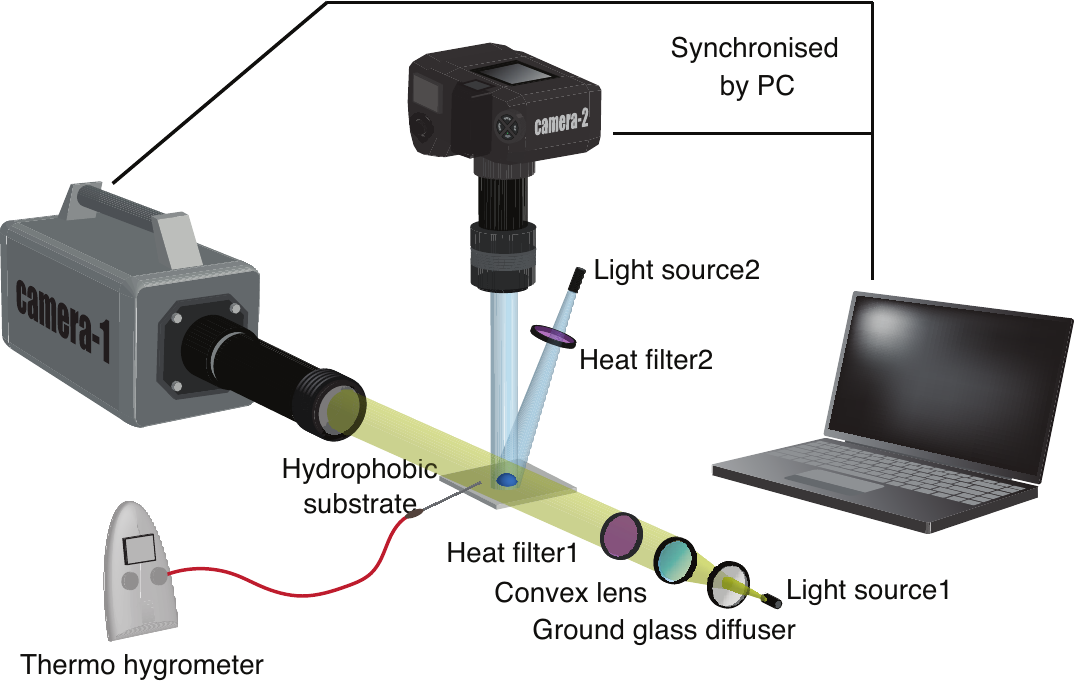}
\caption{Experimental setup showing the evaporation of an Ouzo drop being recorded by two synchronised cameras. A fine needle (not shown here) was used to produce and place the drop on the hydrophobic substrate and then gently moved far away from the experimental region. Heat filter1, a convex lens and a ground glass diffuser were placed in front of light source1 (Schott ACE I) to create a collimated light beam without infrared light. Another heat filter was inserted in the light path of light source 2 (Olympus ILP-1). The ambient temperature and the relative humidity 
%\annotation{see annotation about RH above} 
were determined by a thermo-hygrometer.}
\label{fig:setupsketch}
\end{figure*}

\begin{figure*}[h]
\centering
\includegraphics[width=\textwidth]{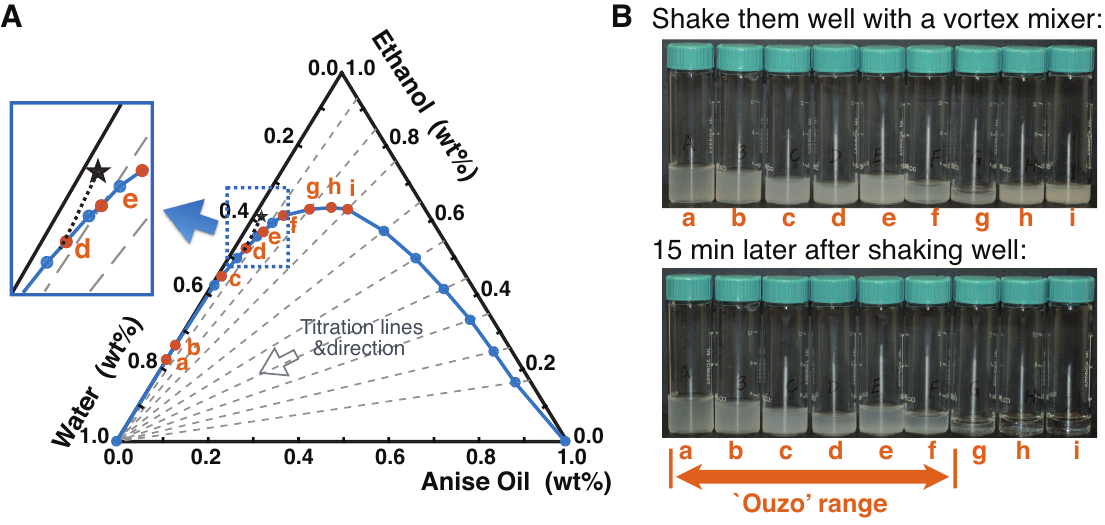}
\caption{\textbf{(A)} shows the ternary diagram of water, ethanol and anise oil. The blue solid line is the measured phase separation curve. The black star and the black dotted line in the magnified figure indicate the initial composition of the Ouzo drop and its path in time according to the numerical simulation. The gray dashed lines show paths of some composition coordinates from the titration experiments; \textbf{(B)} The stability of the macrosuspension for the compositions a-i in the ternary graph were compared. The comparison reveals that the curve along the dots a to f is the boundary of the Ouzo region, i.e. the critical composition at which the Ouzo effect sets in.}
\label{fig:ternarygraph}
\end{figure*}

\begin{figure*}[h]
\centering
\includegraphics[width=\textwidth]{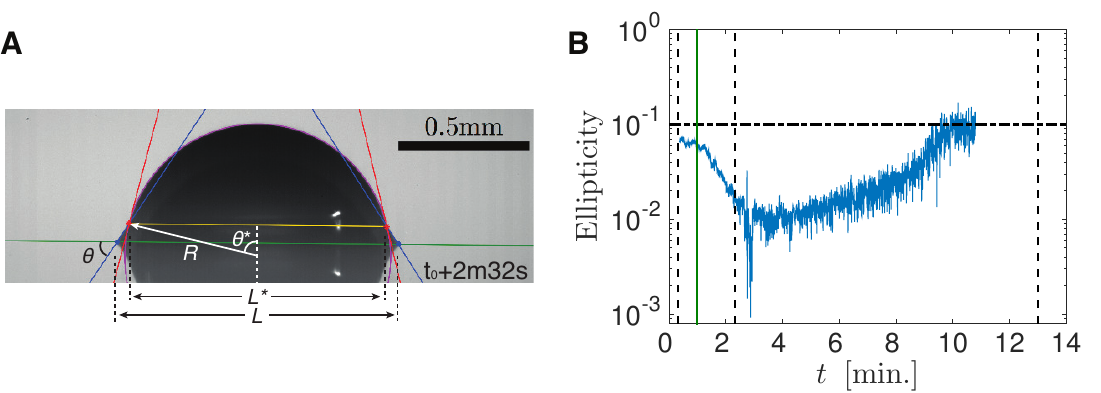}
\caption{Details of the experimental image analysis: \textbf{(A)} A representative raw image displayed with the corresponding results of the image analysis. $\theta$ was estimated by a polynomial fitting; $\theta^*$ was calculated by a spherical cap approximation. \textbf{(B)} The ellipticity, defined as the ratio between the difference between the lengths of the two semi-axes and the radius, is depicted. $\theta^*$ was only calculated for spherical cap approximations with ellipticities less than 10\%. Three black vertical lines are four phases separation moments. Green vertical line indicates the appearance of the oil ring.}
\label{fig:imageanalysis}
\end{figure*}

\begin{figure*}[h]
\centering
\includegraphics[width=\textwidth]{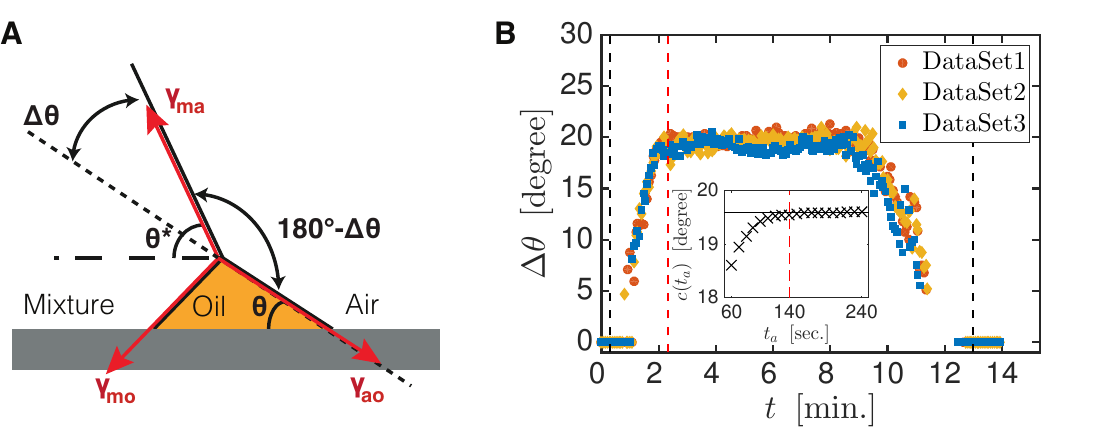}
\caption{\textbf{(A)} Cross-sectional sketch of the oil ring and the equilibrium of the air-mixture-oil contact line. When the mixture predominantly consists of water, the equilibrium is steady and $\Delta\theta$ is constant. \textbf{(B)} shows experimental data of the temporal evolution of $\Delta\theta$. The red vertical dashed line is the separation moment between phases II and III. It is defined as the moment when $\Delta\theta$ starts to be constant. The inserted figure depicts the value $c(t_a)$ fitted over the range ($t_a$,480s) with a constant $c$.} 
\label{fig:def}
\end{figure*}

\begin{figure*}[h]
\centering
\includegraphics[width=0.5\textwidth]{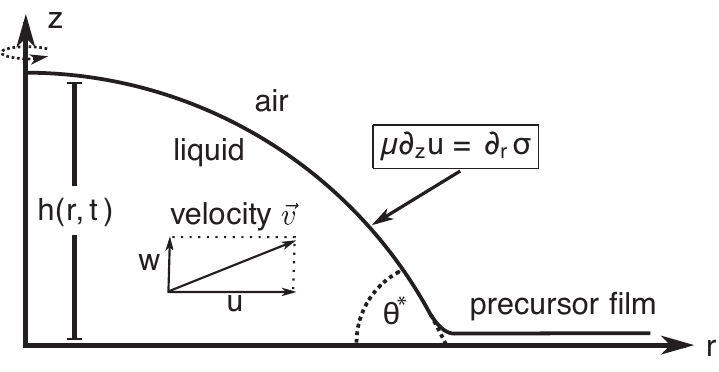}
\caption{Schematic illustration of the model. The shape $h(r,t)$ of the drop is described in axisymmetric cylinder coordinates. Due to different volatilities of the components, a surface tension gradient is induced that drives a Marangoni flow. The moving contact line with contact angle $\theta^*$ is realised by a precursor film.}
\label{fig:model:scheme}
\end{figure*}

\begin{figure*}[h]
\centering
\includegraphics[width=\textwidth]{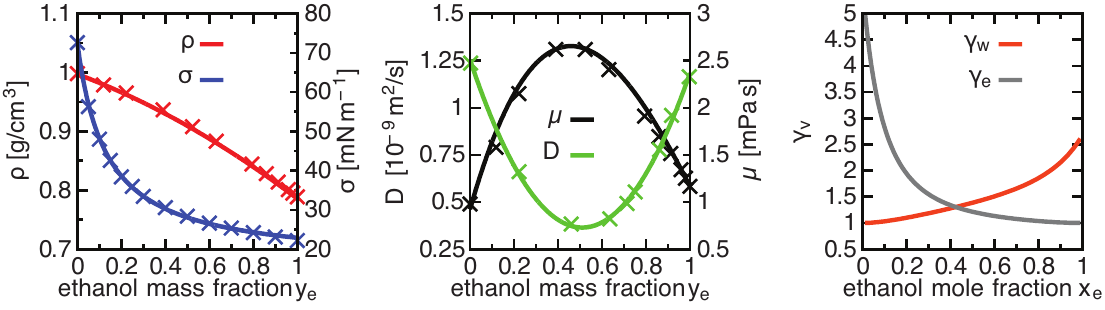}
\caption{Composition-dependence of the physical liquid properties: We have fitted the following experimental data of water-ethanol mixtures for the incorporation into our model: the mass density $\rho$ and the viscosity $\mu$ \cite{Gonzalez2007a}, the surface tension $\sigma$ \cite{Vazquez1995a} and the mutual diffusion coefficient $D$ \cite{Parez2013a}. The activity coefficients $\gamma_{\alpha}$ for the evaporation rate were calculated by \textsl{AIOMFAC} \cite{Zuend2008a,Zuend2011a} (http://www.aiomfac.caltech.edu).}
\label{fig:model:propfits}
\end{figure*}

\setcounter{table}{0}
\makeatletter\renewcommand{\tablename}{Table S}\makeatother

\clearpage
\newpage

\begin{table}
\tiny
\begin{center}
    \caption{Data of the ternary diagram of ethanol-anise-water.}
    \label{tab:ternarygraph}
    \begin{threeparttable}
    \begin{tabular}{p{0.1\linewidth}|
                            >{\centering\arraybackslash}p{0.14\linewidth}
                            >{\centering\arraybackslash}p{0.16\linewidth}
                            >{\centering\arraybackslash}p{0.14\linewidth}|
                            >{\centering\arraybackslash}p{0.07\linewidth}
                            >{\centering\arraybackslash}p{0.07\linewidth}
                            >{\centering\arraybackslash}p{0.07\linewidth}}
      \toprule
      Titration\tnote{1} & \multicolumn2c{Titrant (ethanol-oil mixture)} & Titrate\tnote{2} & \multicolumn3c{Weight ratios\tnote{3}} \\
      \midrule
      No.  & Ethanol(ml) & Anise oil(ml) & Water(ml) & $y_{\text{w}}$(\%) &$y_{\text{e}}$(\%) &$y_{\text{a}}$(\%)\\
      \midrule
      1 & 0 & 0.001 & 6 & 99.98& 0 &0.02 \\
      2  (a)\tnote{4} & 1  & 0.001 & 2.7724 & 73.47& 26.50 &0.03\\
      3  (b) & 1 & 0.002  & 2.2186 & 68.89&31.05&0.06\\
      4 & 1 & 0.01 & 1.0658 & 51.34 &48.17&0.48\\
      5  (c)& 1.2 & 0.02 &  1.1491 & 48.50&50.65&0.84\\
      6 & 1 & 0.03 &  0.7671 & 42.69&55.56&1.67\\
      7  (d)& 1 & 0.04 &  0.6785 & 39.48&58.19&2.33\\
      8 & 1 & 0.05 & 0.5821 & 35.67&61.27&3.06\\
      9  (e)& 1.7 & 0.1 & 0.9211 & 33.85&62.47&3.67\\
      10  & 1.5 & 0.1 &  0.7154 & 30.90&64.78&4.32\\
      11 (f)& 1.2 & 0.1 & 0.5014 & 27.83&66.61&5.55\\
      12 (g)& 0.7 & 0.1 &  0.2261 & 22.03&68.22&9.75\\
      13 (h)& 1 & 0.2 &  0.2563 & 17.60&68.67&13.73\\
      14 (i)& 0.8 & 0.2 &  0.1727 & 14.73&68.22&17.05\\
      15 & 0.7 & 0.3 &  0.1173 & 10.50&62.65&26.85\\
      16 & 0.6 & 0.4 &  0.0842 & 7.77&55.34&36.89\\
      17 & 0.5 & 0.5 &  0.0635 & 5.97&47.01&47.01\\
      18 & 0.4 & 0.6 &  0.0476 & 4.54&38.18&57.27\\
      19 & 0.3 & 0.7 &  0.0404 & 3.88&28.84&67.28\\
      20 & 0.2 & 0.8 &  0.0351 & 3.39&19.32&77.29\\
      21 & 0 & 1 & 0.0041 & 0.41&0&99.59\\
      \bottomrule
    \end{tabular}
    \begin{tablenotes}
        \item[1] The titration was conducted at a temperature of around \SI{22}{\celsius}.
    	\item[2] Aliquot was 0.0015ml, which was the minimum volume of the water droplet created by the pipette needle during titration.  
	\item[3] Density of anise oil at \SI{22}{\celsius} was measured as \SI{0.989}{\gram\per\milli\litre}. Water and ethanol density at \SI{22}{\celsius} was obtained from a handbook \cite{lide2004crc} by linear interpolation.
	\item[4] Corresponding to the labels in the ternary diagram (cf. Fig. S2A).
    \end{tablenotes}
    \end{threeparttable}
    \end{center}
\end{table}

\newpage
\section{Captions for Videos S1 to S6}
\subsection{Video S1} Experimental top-view recording of an evaporating Ouzo drop (synchronised with Video S2 in a same experiment). The initial volume of the drop is \SI{0.7}{\micro\liter} with an initial composition of  \SI{37.24}{\percent} water, \SI{61.06}{\percent} ethanol and \SI{1.70}{\percent}  anise oil (a mixture we refer to as 'Ouzo') in terms of weight fractions. The experiment was performed in the experimental setup as shown in Fig. S1.

\subsection{Movie S2} Experimental side-view recording of an evaporating Ouzo drop (synchronised with Video S1 in a same experiment). 

\subsection{Movie S3} Experimental bottom-view recording of an evaporating \SI{0.7}{\micro\liter} Ouzo drop of the same composition as in Videos S1 and S2. The experiment was performed with an inverted microscope (Olympus GX51, 20$\times$ magnification).  

\subsection{Movie S4} Animation of an evaporating Ouzo drop with an initial composition of  \SI{37.24}{\percent} water, \SI{61.06}{\percent} ethanol and \SI{1.70}{\percent}  tans-Anethole oil, displaying the whole evaporating process in a confocal view (20$\times$ magnification). The movie was created by a confocal microscope system.

\subsection{Movie S5} Animation of an evaporating Ouzo drop with same composition as Video S1, displaying the three-dimensional dynamic motion of oil droplets in the contact line region at early times of phase II (20$\times$ magnification), created by a confocal microscope system. 

\subsection{Movie S6} Numerical simulation of an evaporating Ouzo drop. The detailed description of the model can be found in Materials and Methods and Supporting Information.

\end{document}